\documentclass[
preprint,
amsmath,amssymb,
aps,
prb,
floatfix,
superscriptaddress,
]{revtex4-2}
\usepackage{float}

\usepackage{graphicx}
\usepackage{dcolumn}
\usepackage{bm}
\usepackage{hyperref}

\usepackage{multirow}
\usepackage{graphicx}
\usepackage{xcolor}
\usepackage{cancel}

\begin{document}

\title{Anomalous Tip-Sample Distance Behavior on the Tip-Enhanced Raman Spectroscopy of Graphene in Ambient Conditions}

\author{André G. Pereira}
\affiliation{Departamento de Física, Universidade Federal de Minas Gerais, Belo Horizonte, Minas Gerais, 30123-970 Brazil} 
\affiliation{FabNS, Parque Tecnológico de Belo Horizonte-BHTec, Belo Horizonte, Brazil} 

\author{Raul Corrêa}
\affiliation{Departamento de Física, Universidade Federal de Minas Gerais, Belo Horizonte, Minas Gerais, 30123-970 Brazil} 
\affiliation{Instituto D'Or de Pesquisa e Ensino/Pioneer Science Initiative, Rio de Janeiro, RJ 22281-010, Brazil} 

\author{Bianca Carneiro}
\affiliation{Departamento de Física, Universidade Federal de Minas Gerais, Belo Horizonte, Minas Gerais, 30123-970 Brazil} 

\author{Cassiano Rabelo}
\affiliation{FabNS, Parque Tecnológico de Belo Horizonte-BHTec, Belo Horizonte, Brazil} 

\author{Thiago L. Vasconcelos}
\affiliation{Instituto Nacional de Metrologia, Qualidade e Tecnologia, Duque de Caxias, RJ 25250-020, Brazil} 

\author{Vitor Monken}
\affiliation{FabNS, Parque Tecnológico de Belo Horizonte-BHTec, Belo Horizonte, Brazil} 
\affiliation{Programa de Pós-Graduação em Inovação Tecnológica e Propriedade Intelectual, Universidade Federal de Minas Gerais, Belo Horizonte, Brazil} 

\author{Luiz Gustavo Cançado}
\author{Ado Jorio}
 \email{adojorio@fisica.ufmg.br}

\affiliation{Departamento de Física, Universidade Federal de Minas Gerais, Belo Horizonte, Minas Gerais, 30123-970 Brazil} 

\date{\today}

\begin{abstract}
Tip-Enhanced Raman Spectroscopy (TERS) combines Raman spectroscopy with scanning probe microscopy to overcome the spatial resolution limitation imposed by light diffraction, offering a primary optical technique for the comprehensive study of two-dimensional (2D) materials. In this work, we investigate an anomalous decay profile of the TERS intensity of the graphene 2D band as the tip-sample separation changes, observations enabled by high TERS efficiency and accuracy in tip-approach and tip-retract procedures. The anomalous results can be properly described by the addition of an ad hoc deformation to the effective tip-sample distance, rationalized here as due to the presence of a liquid meniscus formed via capillary forces.
\end{abstract}

\maketitle

Raman spectroscopy has been widely applied in the study of two-dimensional (2D) materials, such as graphene, due to its ability to nondestructively probe the structural, electronic, and chemical properties of the materials~\cite{jorio2010}. When it comes to nanoscale properties, the diffraction limit poses a significant limitation to conventional Raman spectroscopy, inherent to far-field optical techniques~\cite{Abbe1873, Rayleigh01101879}. To overcome this, Tip-Enhanced Raman Spectroscopy (TERS), a type of scanning near-field optical microscopy (SNOM), is often implemented in which the resolution is generally given by the tip radius instead of Abbe's limit~\cite{zenobi2000,hayazawa2000metallized,anderson2000locally,novotny2012principles}.

In this work, we reveal an anomalous decay profile in the TERS signal as the tip-sample separation changes in ambient conditions. The results will be rationalized as due to the formation of a liquid meniscus. To properly describe our results, the available TERS intensity theoretical model is expanded to incorporate this observed effect, enabling the fitting of the experimental data with accuracy.

The TERS experiments were performed in ambient conditions on a sample of graphene mechanically exfoliated~\cite{graphene} from National Graphite source material. The graphene flake was deposited onto a Knittel coverslip using the pickup transfer method~\cite{Andreij2021}. Although airborne contaminant adsorption renders the graphene layer hydrophobic~\cite{MARTINEZMARTIN201333, akaishi}, the literature claims that a layered structure of water molecules persists on the graphene surface~\cite{akaishi, LUNA2000393, Freund1999327}. Prior works have shown that when a tip is placed very close to the sample, a meniscus can be formed~\cite{seabron, Ohlberg2021}.

The measurements were carried out on an AFM-TERS setup, the Porto laboratory prototype~\cite{Rabelo2019}. This system is equipped with an oil immersion objective lens ($\text{NA} = 1.4$). The TERS tips used are the Plasmon-Tunable Tip Pyramid (PTTP)~\cite{Thiago2018}, which are mounted on one prong of a quartz tuning fork. The excitation source is a radially polarized HeNe laser ($\lambda = 632.8 \, \text{nm}$), with power 200 or 700 $\mu\text{W}$, depending on the experiment. The back-scattered signal is then collected by the same objective lens used for excitation and directed to an Andor Shamrock SR-303i spectrometer for analysis. Data processing was carried out using the PortoFlow software by FabNS (v1.21).

\begin{figure*}
    \centering
    \includegraphics[width=1\textwidth]{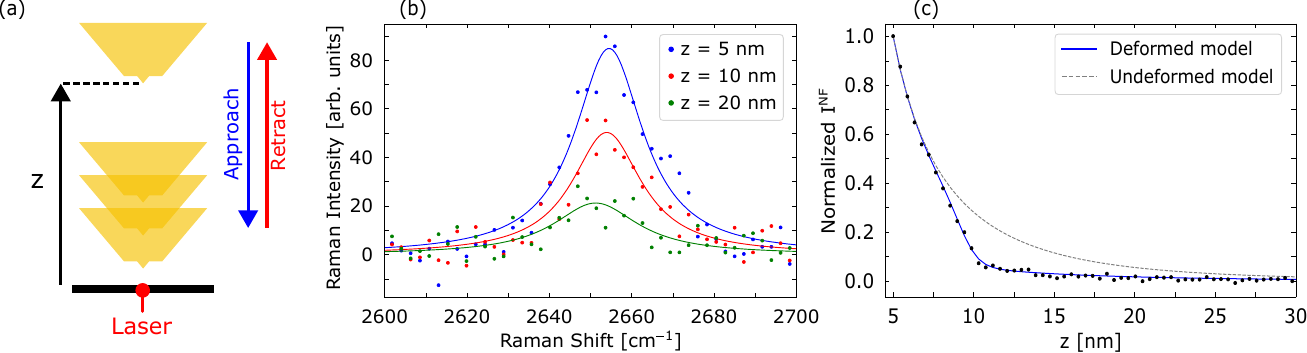}
    \caption{(a) Schematic diagram of the approach and retract experiments, where $z$ defines the tip-sample distance. (b) Raman spectrum of the 2D band of graphene at three different $z$ values (see legend), where the closest distance is assumed as $z_0=5$\,nm~\cite{novotny2012principles}. (c) Normalized TERS near-field intensity $I^{NF}(z)$ of the 2D band (data points), overlaid with the fitting curves derived from the two TERS models discussed in this work, on a tip-retraction experiment.}
    \label{fig:diagram}
\end{figure*}

Figure~\ref{fig:diagram}a shows a schematic diagram of tip-sample separation during the approach and retraction cycles. In this experiment, the tip approached until a frequency shift of $0.9 \, \text{Hz}$ (the setpoint of the system) was reached from the central frequency, and then it was retracted. Simultaneously, the Raman signal was collected with an excitation laser power of $700 \, \mu \text{W}$.

The Raman spectrum of graphene is characterized by the prominent G band ($\sim 1580$\,cm$^{-1}$) and the higher-order 2D band, observed around $2650 \, \text{cm}^{-1}$~\cite{jorio2010}. Due to the higher intensity of the 2D peak~\cite{ferrari2006raman}, we focus on the 2D band in this work, although the G peak exhibits a similar behavior.

Figure~\ref{fig:diagram}b shows the 2D band of graphene fitted with a Lorentzian function at three different tip-sample distances during a retraction experiment. From now on, the tip-sample distance will be denoted by $z$. Figure~\ref{fig:diagram}c presents the normalized TERS near-field intensity decay [$I^{NF}(z)$] as the tip moves away from the sample: $I^{NF}(z) = I^{TERS}(z) - I^{TERS}(z=30)$, where $I^{TERS}(z=30)$ is equivalent to the far-field (without the tip) intensity. Also presented in this figure are two fitting models, which will be discussed later. For now, it is important to note that the undeformed TERS model (gray dashed line~\cite{cancado}) is not capable of fitting the experimental data, highlighting the necessity of a deformation correction to achieve a satisfactory fit (blue line).

A total of 54 tip-retraction experiments were conducted using different PTTP tips and sample types across two independent laboratories, with air humidity between 40\% and 55\%. From those, 44 of the experimentally obtained curves showed a significant discrepancy with the expected decay (undeformed model~\cite{cancado}) marked by a discontinuity at the range $z = 8-18$\,nm, sometimes more pronounced than that presented in Fig.~\ref{fig:diagram}c. Representative decay curves and a detailed statistical analysis of the full experimental set are available in the supplementary material.

The resulting discrepancy between the experimental results and the undeformed TERS model in Fig.~\ref{fig:diagram}c will be rationalized by incorporating the effect of a liquid meniscus on the graphene surface. To understand this effect, it is important to consider that our AFM works in constant drive and shear force configuration~\cite{hartschuh2003high}. Its control is based on a phase-locked loop (PLL) that monitors several parameters of the tuning fork oscillation, such as the phase, amplitude, and frequency variation. We focus here on the amplitude to support the hypothesis of meniscus formation, and information about frequency variation can be found in the supplementary material. In constant drive mode, as the name implies, the tuning fork is excited with a fixed amplitude. Therefore, variations in the amplitude, frequency, and phase of the tuning fork oscillation reflect changes in the tip's interaction with the surrounding medium.

\begin{figure}[!htb]
    \centering
    \includegraphics[width=0.5\columnwidth]{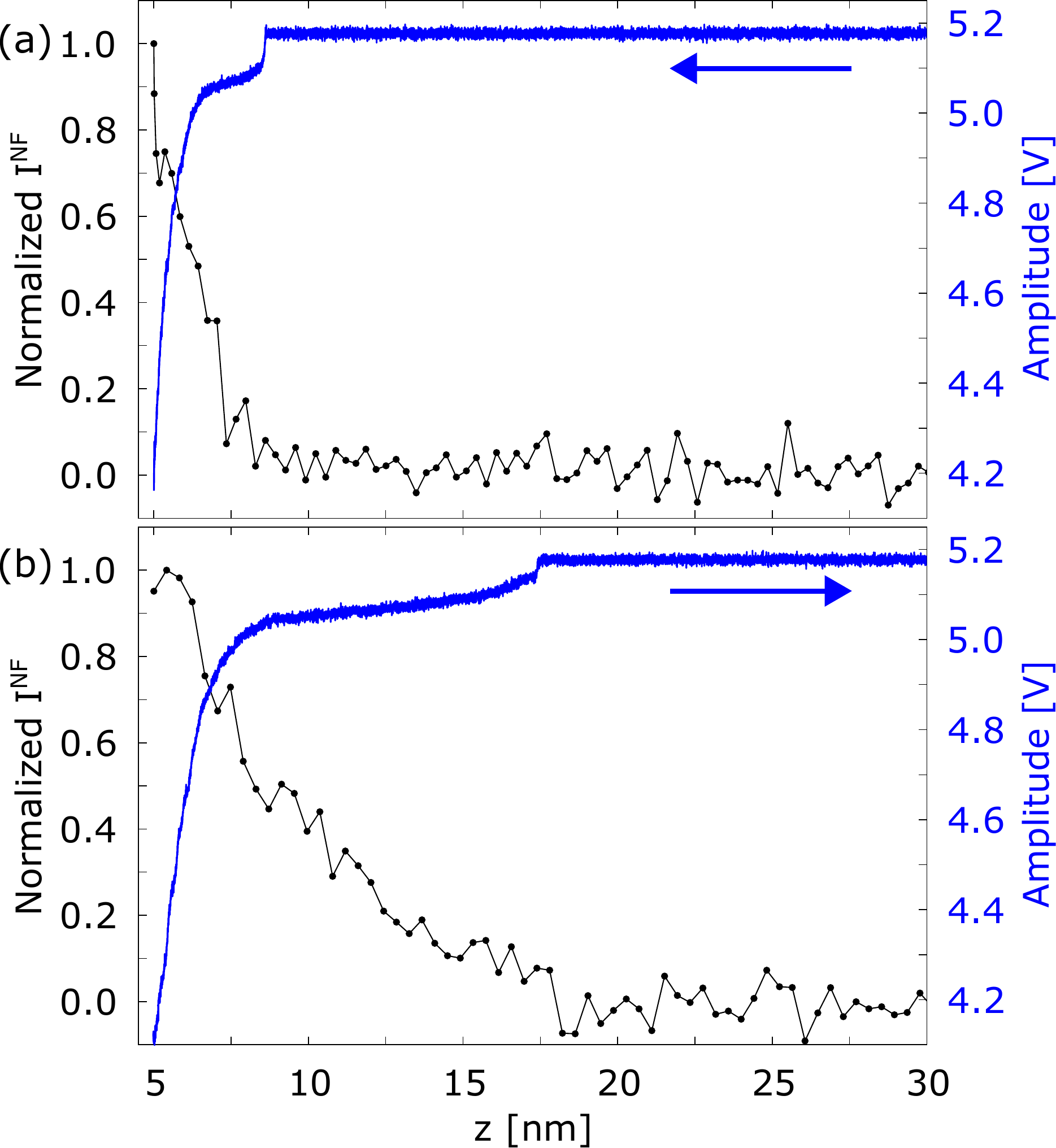}
    \caption{Normalized TERS near-field intensity decay $I^{NF}(z)$ alongside the corresponding tuning fork oscillation amplitude for the approach (a) and retraction (b) experiments. The arrows indicate the direction of the tip movement.}
    \label{fig:amp}
\end{figure}

\begin{figure*}
    \centering
    \includegraphics[width=1\textwidth]{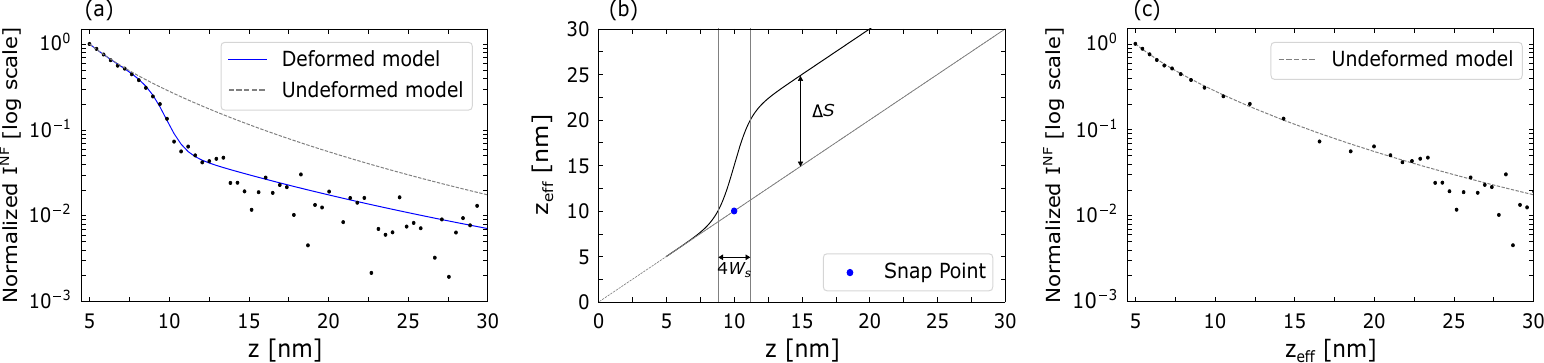}
    \caption{(a) Normalized TERS intensity decay $I^{NF}(z)$ of the 2D band (black circles) and the fitted curve from both (deformed and undeformed) models as a function of the mechanical $z$ separation (same as Fig.~\ref{fig:diagram}c, here in log scale). (b) Illustration of the deformation function $D(z)$, where $\Delta S$ (snap amplitude) represents the magnitude of the offset and $W_s$ (snap width) controls the transition range. (c) Normalized TERS intensity decay of the 2D band (black circles) plotted versus the effective separation $z_{\text{eff}}$. The fitted theoretical curve (gray dashed line) shown in (c) is the undeformed model~\cite{cancado} plotted against the undeformed $z$ values.}
    \label{fig:model}
\end{figure*}

Figure~\ref{fig:amp} depicts the evolution of the TERS near-field intensity $I^{NF}(z)$ along with the tuning fork oscillation amplitude in two subsequent experiments, a tip approach (a) and a tip retract (b). Two characteristics are clear:

(1) A discontinuity is present in the tuning fork amplitude oscillation, occurring at a different $z$ value in the approach and retract experiment.

From Figure~\ref{fig:amp}, a steep and pronounced decay in amplitude is observed for the approach curve at a tip-sample separation of approximately 7\,nm. In contrast, during retraction, this steep decay occurs at approximately 17.5 \,nm, and is less prominent than the approach step. This behavior, including both the difference in the tip-sample distance where the decay begins and the greater magnitude of the amplitude step during approach, is consistent with previous works~\cite{sudersan, Freund1999327, HERMINGHAUS1997211}. This distance difference is explained by the meniscus stretching before its rupture during retraction, whereas, in the approach, the tip must come close to the sample to engage with the thin water film. As the tip comes close enough to the water film, the meniscus is rapidly formed, causing the sudden and large amplitude step (snap-in)~\cite{pablo}. On the other hand, as the tip is retracted from the sample, the meniscus stretches, and the capillary contact perimeter recedes. This gradual reduction causes the step-out decay to be less prominent than the snap-in. It is important to stress that the term {\it snap} here is related to the tuning fork oscillation amplitude, and the mechanical reason is not identified.

(2) The TERS intensity of the 2D band is correlated with the tuning fork oscillation amplitude, increasing precisely below the snap point.

Figure~\ref{fig:model}a shows the normalized $I^{NF}(z)$ (same data as in Fig.~\ref{fig:diagram}c, but here in logarithmic scale). The gray dashed line represents the best fit curve obtained using the TERS theory (typical fitting parameters: tip radius $r_{\text{tip}} = 8 \, \text{nm}$, tip enhancement factor $f_e = 6$, and TERS coherence length $L_c = 10 \, \text{nm}$, see Ref.~\cite{cancado} for details). Evidently, this initial curve does not provide a satisfactory fit to the experimental data due to the discontinuity decay at $z\sim 10$\,nm.

To account for the meniscus effect, we incorporate a length-scale factor into the analysis through an ad hoc smooth deformation function
$D(z)$,
\begin{equation}\label{deformation}
    D(z) =  \left(\frac{1}{1+ \text{exp}\left(\frac{z - z_{snap}}{W_s}\right) }\right)  \times \Delta S,
\end{equation}
that acts on the $z$ separation. The final effective separation ($z_{eff}$) is the sum of the original $z$ separation and the deformation $D(z)$ ($z_{eff} = z + D(z)$). The TERS near-field intensity is then calculated based on this $z_{eff}$ ($I^{NF}(z_{eff})$) and plotted as a function of the original $z$ separation. This effectively deforms the curve around the snap point ($z_{\text{snap}}$), where the magnitude of the deformation is controlled by $\Delta S$ and the transition range is determined by the width $W_s$.

Figure~\ref{fig:model}b shows the deformation function $D(z)$ (black line) required to fit the experimental data. The
separation $z$ before the snap point (while the tip is still inside the meniscus) is taken as a reference. We then use the
$D(z)$ function to calibrate the $z$ values after the snap point, making the system behave as if the tip were farther away from the sample once it was outside the meniscus.

By replacing $z$ by $z_{eff}$ in the TERS theory~\cite{cancado} the final fitted curve (blue line) shown in Figures~\ref{fig:diagram}a and \ref{fig:model}a is obtained, showing excellent agreement with the experimental data (optimized parameters present in Table~\ref{tab:parameter}).

Conversely, when the experimental data is re-plotted against the effective optical separation $z_{\text{eff}}$ (Figure~\ref{fig:model}c), the resulting data exhibit the smooth decay expected in the undeformed theoretical model.

\begin{table}
\caption{\label{tab:parameter}Deformation function (Eq.(1)) parameters to fit the data in Fig.~\ref{fig:model}. }
\begin{ruledtabular}
\begin{tabular}{cccc}
&Parameter & Value&\\
\hline
&Snap point & 10 nm& \\
&$\Delta S$ & 10 nm& \\
&$W_s$ & 0.6 nm&\\
\end{tabular}
\end{ruledtabular}
\end{table}

In conclusion, we presented the anomalous decay profile of the TERS intensity of the graphene 2D band as the tip-sample separation increases. To explain this behavior, we argue that a liquid meniscus forms when the TERS tip approaches the sample, as evidenced by the hysteresis observed in the AFM amplitude. Then, we expand the previous TERS intensity theory to account for this effect by introducing an ad hoc deformation function for the effective tip-sample separation ($z_{\text{eff}}$), resulting in a satisfactory fit to the experimental data.

If we consider the forces involved, it is unlikely that $z_{\text{eff}}$ is the real measure of the tip-sample distance, because the meniscus is not expected to change the tip position so dramatically. One possible explanation for the meniscus-induced deformation is that the meniscus acts like an optical waveguide. The light scattered by both the tip and the sample propagates in all directions. In the absence of the meniscus, part of this scattered light would be lost in the medium and would not reach the detector. However, when the meniscus is formed, its curved surface can internally reflect some of this scattered light and, by doing so, concentrate the light between the tip and the sample, as discussed for the case of scanning microwave impedance microscopy (sMIM)~\cite{Ohlberg2021}. Additionally, because the refractive index of the objective oil and of the cover slip ($n \sim 1.5$) is close to that of water ($n\sim 1.3$), the enhanced signal is collected more efficiently in the presence of the meniscus, reducing light reflection at the sample surface.

No Raman peaks associated with water were found in the spectra. This absence may be due to the low scattering volume of the nanoscale water layer. However, this is intriguing because experiments performed on MoSe$_2$ with a similar equipment evidenced the Raman signal from small amounts of contamination in the surface~\cite{jane2025}. Experiments performed in vacuum may shed light on these results.

\section*{SUPPLEMENTARY MATERIAL}
See the supplementary material for the classification and quantitative analysis of the TERS decay regimes and the analysis of the AFM frequency variation response.

\section*{ACKNOWLEDGMENTS}
The authors acknowledge professor Bernardo Neves for the valuable discussion.
Financial support by CNPq (Brazil) (408697/2022-9, 351472/2024-0, 421469/2023-4) and Fapemig (Brazil) (APQ-04852-23, APQ-01860-22, RED-00081-23).

\section*{AUTHOR DECLARATIONS}
\subsection*{Conflict of Interest}
The authors have no conflicts to disclose.

\section*{DATA AVAILABILITY}
The data that support the findings of this study are available from the corresponding author upon reasonable request.


%

\end{document}